\documentclass[doublecol]{epl2} 

\usepackage{mathtools}

\title{Stabilizing quasicrystals composed of patchy colloids by narrowing the patch width}

\author{A. Gemeinhardt\inst{1} \and M. Martinsons\inst{1} \and M. Schmiedeberg\inst{1}}
\shortauthor{A. Gemeinhardt \etal}

\institute{                    
  \inst{1} Institut f\"ur Theoretische Physik I, Friedrich-Alexander-Universit\"at Erlangen-N\"urnberg, Staudtstra{\ss}e 7, 91058 Erlangen, Germany
}

\abstract{
We explore the behavior of two-dimensional patchy colloidal particles with 8 or 10 symmetrically arranged patches by employing Monte-Carlo simulations. The particles interact according to an isotropic pair potential that possesses only one typical length. The patches lead to additional attractions that are anisotropic and depend on the relative orientation of two neighboring particles. We investigate the assembled structures with a special interest in quasicrystals. We found that the patch width is of great importance: Only in case of narrow patch widths we are able to observe metastable octagonal and decagonal quasicrystals, while dodecagonal quasicrystals can also occur for broad patches. These results are important to understand the role of interactions with preferred binding angles in order to obtain quasicrystals. Our findings suggest that in case of sharp binding angles, as they occur in metallic alloys, octagonal and decagonal symmetries might be observed more often than in systems with less sharp binding angles as it is the case in soft matter systems where dodecagonal quasicrystals dominate.
}

\begin{document}

\maketitle

This is the version of the article as submitted by an author to EPL. The Version of Record is available online at https://doi.org/10.1209/0295-5075/126/38001.\\
Citation of the final version: A. Gemeinhardt et al 2019 EPL {\bf 126} 38001\\

\section{Introduction}

Quasicrystals, i.e. well-ordered structures without periodic translational symmetry \cite{Shechtman,Levine}, can possess any rotational symmetry including non-crystallographic ones. In two dimensions, quasicrystals with 5-, 8-, 10-, or 12-fold rotational symmetry possess at least two incommensurate length scales as well as two additional degrees of freedom termed phasons \cite{Levine2,Socolar,Kromer,Sandbrink,Martinsons,Hielscher}. They seem to build more likely than quasicrystals with other symmetries \cite{Mikhael,Schmiedeberg} that would possess even more degrees of freedom (see, e.g. \cite{Martinsons}).

In experiments, most quasicrystals have been found in metallic alloys and provide icosahedral or -- less often -- 8-, 10- or 12-fold symmetry (see, e.g. \cite{Macia,Steurer}). Soft matter quasicrystals have been observed as well \cite{Zeng,Zeng2,Takano,Hayashida,Fischer,Dotera} and most of them exhibit dodecagonal symmetry \cite{Dotera}. 

In simulations, quasicrystals can be stabilized, e.g. by one-component particles interacting with an isotropic pair potential with at least two incommensurate lengths \cite{Denton,Engel,Engel2,Barkan,Achim,Barkan2,Dotera2,Savitz}. Another approach are patchy colloids (see, e.g. \cite{Glotzer}), i.e. particles with attractive regions at the surface. The number, arrangement and width of such patches is tunable. In recent simulations \cite{vdLinden,Reinhardt,Gemeinhardt} many patchy colloids -- even in case of 5 symmetrically arranged patches per particle -- assembled into quasicrystals with dodecagonal symmetry, which has been the only quasicrystalline symmetry achieved in systems with patchy colloids. Let us note that patchy colloids have also been realized in experiments to study the self-assembly of complex structures (see, e.g. \cite{Bianchi,Pawar}).

In systems of patchy colloids the ordering preferred by the isotropic part of the potential (with only one length scale) competes with structures that possess the binding angles preferred by the patches. Competitions between different symmetries are known to lead to interesting new phenomena, like the formation of Archimedean-tiling phases \cite{Patrykiejew,Schmiedeberg3,Mikhael2,Schmiedeberg2}, rhombic phases \cite{Neuhaus}, or new types of growth behavior \cite{Neuhaus2,Neuhaus3}. Here we report on similar intermediate orderings of patchy colloids. However, our main goal is to obtain quasicrystalline structures with octagonal and decagonal symmetry. We will focus on the influence of different patch widths on the final arrangements. 

\section{Method}

\begin{figure}[htb]
\centering
\includegraphics[scale=1]{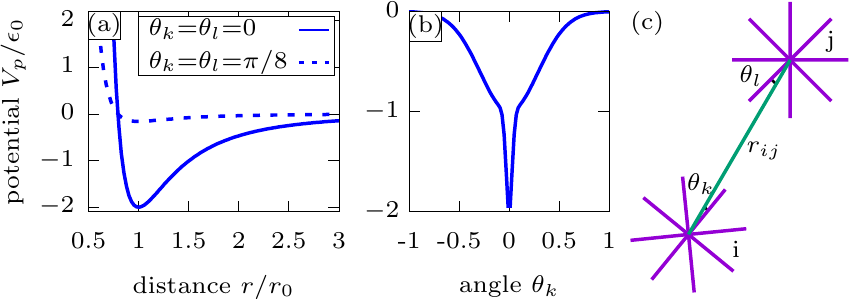} 
\caption{(a) and (b) Potential $V_{p}(r,\theta_{k},\theta_{l})$ with patch widths $\sigma_{1}=0.02$ and $\sigma_{2}=0.29$ and Lennard-Jones exponent $n=3$. The potential stabilizes octagonal quasicrystals in case of particle decorated with 8 patches (see next Section). (a) depicts the potential as a function of the pair distance $r/r_{0}$. The minimum is deepest if the patches are oriented towards each other (solid line) and lowest for opposite patches (dashed line). (b) illustrates the dependence of the potential on the binding angle $\theta_{k}$ at fixed $\theta_{l}=0$ and $r=r_{0}$. Once the narrow patches are oriented towards each other, i.e. the potential minimum is reached, the broad patches do not have an influence on the structure anymore. (c) Particles with 8 symmetrically arranged patches.}
\label{fig:potential}
\end{figure}

We consider an interaction potential that is similar to the ones previously used to investigate self-assembly or growth processes \cite{Doye,Noya,Wilber,Doppelbauer,Noya2,Williamson,Doppelbauer2,vdLinden,Reinhardt,Gemeinhardt}. The interaction $V_{p}(r,\theta_{k},\theta_{l})$ between two particles at a distance $r$ is composed of an isotropic Lennard-Jones-like pair potential $V_{LJ}(r) = \epsilon \left[\left(r_{0}/r\right)^{2n} - 2\left(r_{0}/r \right)^{n} \right]$ multiplied with an anisotropic angular term $V_{a}(\theta_{k}, \theta_{l})$, i.e. $V_{p}(r,\theta_{k},\theta_{l})= V_{LJ}(r) \cdot V_{a}(\theta_{k}, \theta_{l})$ for $r > r_{0}$ and $V_{p}(r,\theta_{k},\theta_{l})= V_{LJ}(r)$ otherwise. Note that the patches only act starting at the surface, i.e. for $r > r_{0}$. The  Lennard-Jones-like part possesses one local minimum at $r=r_{0}$ and the exponent $n$ determines the width of the minimum. Note that we consider a radial Lennard-Jones-like part that is broader than the normal Lennard-Jones potential with $n=6$, because for $n=6$ the suppression of nearest neighbor distances that differ from the one favored length scale $r_{0}$ seems to be too strong to allow the stabilization of quasicrystals different from dodecagonal structures, where all triangles and squares that occur have the side length $r_0$.

The angular term $V_{a}(\theta_{k}, \theta_{l})$ models the attractive patches. We use
$V_{a}(\theta_{k}, \theta_{l}) = e^{-(\theta_{k}^{2} + \theta_{l}^{2})/\left(2 \sigma_{1}^{2}\right)}+e^{-(\theta_{k}^{2} + \theta_{l}^{2})/\left(2 \sigma_{2}^{2}\right)}$,
where $\theta_{k}$ and respectively $\theta_{l}$ denote the angles between the nearest patch of particle $i$ and respectively $j$ to the bond $r_{ij}$ between particle $i$ and $j$ (see fig. \ref{fig:potential}(c)). In fig. \ref{fig:potential} (a) and (b) we illustrate the complete potential as a function of the distance and respectively the angle between the patches. As a slight modification from recent works \cite{Doye,Noya,Wilber,Doppelbauer,Noya2,Williamson,Doppelbauer2,vdLinden,Reinhardt,Gemeinhardt} our angular part consists of two terms with two patch widths $\sigma_{1}$ and $\sigma_{2}$. Thus, we can model a narrow patch width $\sigma_{1}$, while the superposed attraction with a broad patch width $\sigma_{2}$ ensures that the patches still find each other within a reasonable time. Note that the broad patches do not influence the final structure. We will see in the next section that narrow patches (i.e. small $\sigma_{1}$) are required to stabilize octagonal and decagonal quasicrystals. The potentials are truncated and shifted appropriately at $3r_0$.

We model two-dimensional systems of $N$ particles by employing Metropolis Monte-Carlo simulations with periodic boundary conditions in the $NVT$-ensemble. If not stated otherwise conventional displacement moves and rotation moves are proposed with the same probability. Simulations are started either with random positions of the particles or with particles placed on an ideal lattice. An ideal decagonal arrangement is obtained from the maxima of the interference pattern of five laser beams \cite{Gorkhali,Schmiedeberg4,Schmiedeberg}. An ideal octagonal lattice is received from substitution rules \cite{Gruenbaum}. In case of random initial positions the orientations of the particles are also chosen randomly. In the ideal lattices the patches are oriented towards each other. 

We use several analysis tools to characterize the order of the structures that the patchy colloids adopt and to identify quasicrystals. For the analysis of the rotational symmetry we calculate the structure factor $S(\bm{q}) = \frac{1}{N} \sum_{j} \sum_{k}\mathrm{exp}[2\pi \mathrm{i} \bm{q}(\bm{r}_{j} - \bm{r}_{k})]$. Furthermore, the bond-orientational order parameter for a particle $j$ is $\psi_{m}(\bm{r}_{j}) = \frac{1}{N_{k}} \sum \mathrm{exp}(\mathrm{i}m\theta_{jk})$ and characterizes the structures according to a given rotational symmetry $m$, e.g. $m=10$ to test for decagonal symmetry. The sum runs over all neighboring particles $k$ with a distance of the short length scale. $\theta_{jk}$ denotes the angle between the bond from particle $j$ to $k$ and an arbitrary direction. In addition, the angular distribution function $g(\phi)$ counts how often bond angles between nearest neighbor particles occur.

\section{Results}

\begin{figure*}[htb]
\centering
\includegraphics[scale=1]{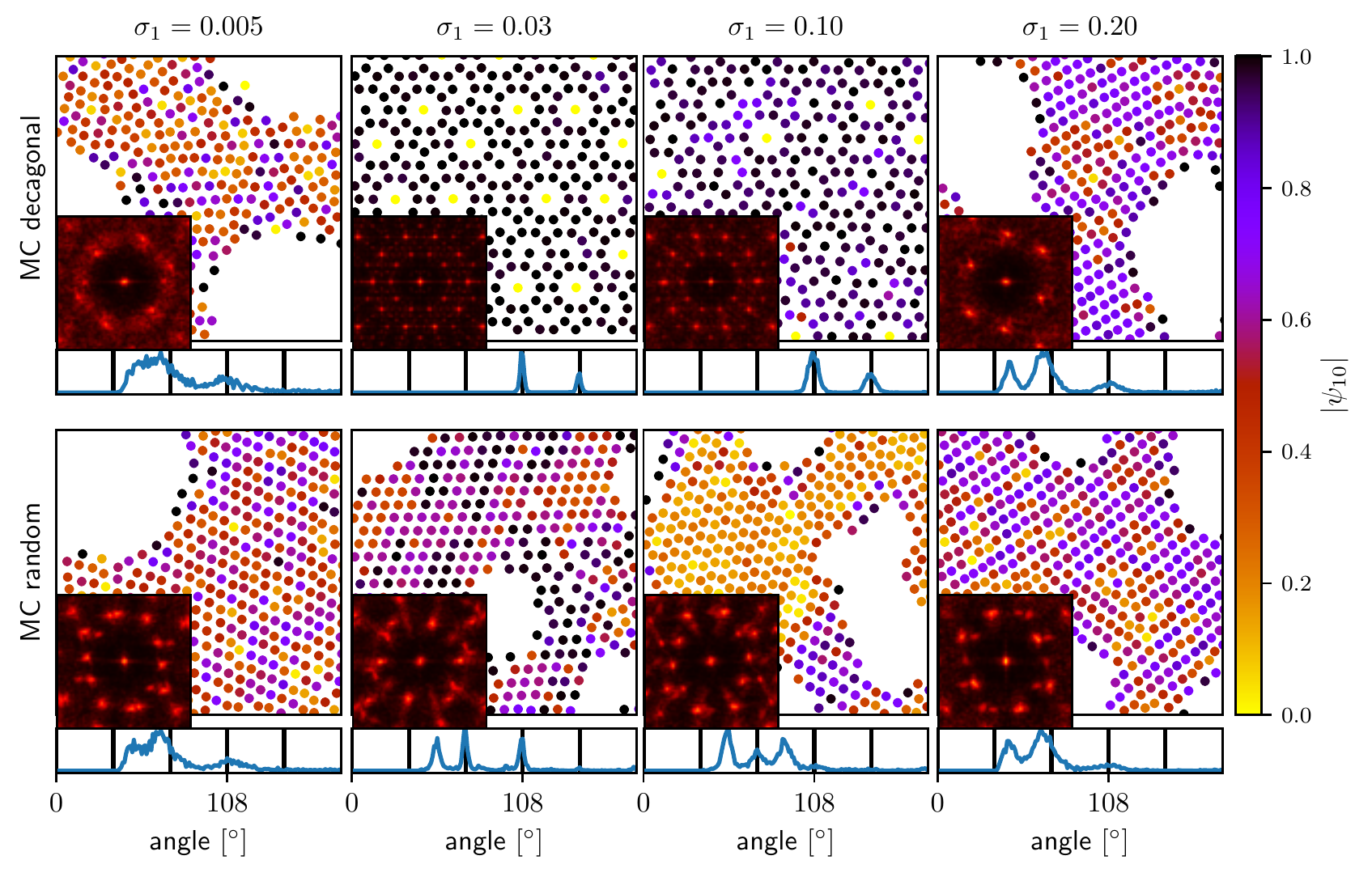} 
\caption{Configurations of $N=644$ particles obtained from simulations with an initial decagonal quasicrystal (upper row) and an initial random configuration (lower row). We vary the patch width $\sigma_{1}=0.005$ (first column), $\sigma_{1}=0.03$ (second column), $\sigma_{1}=0.10$ (third column) and $\sigma_{1}=0.20$ (fourth column). Further potential parameters read $\sigma_{2}=0.23,\, n=2,\, r_{0}=d_{0}$ and we model colloids furnished with 10 patches. The color code illustrates the bond-orientational order parameter $|\psi_{10}|$. The insets depict the corresponding structure factors. The graphs below the configurations show the corresponding angular distribution functions. The black lines serve as guide to the eye and indicate ideal decagonal bond angles, i.e. $2\pi j/10, j=1,\ldots,10$. All simulations were performed at $T=0.2\epsilon$.}
\label{fig:overview10}
\end{figure*}

We first model particles with 10 symmetrically arranged patches with the aim to stabilize decagonal quasicrystals. Ideal decagonal structures possess two characteristic length scales $d_{0}$ and $d_{1}=\tau d_{0} \approx 1.618 d_{0}$, where $\tau$ denotes the golden mean. In our simulations, however, we can only support one length by the potential minimum. From previous works \cite{vdLinden,Reinhardt,Gemeinhardt} we know that dodecagonal quasicrystals can be stabilized with patchy colloids when the short length is supported. Therefore, here we also apply $r_{0}=d_{0}$. We choose a small exponent $n=2$ which causes a broad potential minimum. Thus, the low gradient reduces the energy of the long length scale. For the attraction that is used to support the fast finding of bonds, we employ a broad width $\sigma_{2}=0.23$. $\sigma_{2}$ is chosen in a way that the surface area covered by patches is similar like in previous works \cite{vdLinden,Reinhardt,Gemeinhardt}. To model the actual patches we use a different width $\sigma_{1}$ that is varied. The temperature reads $T=0.2\epsilon$ which is below the melting temperature but large enough to allow for sufficient mobility of the particles. The density of a perfect decagonal tiling is chosen, i.e. $\rho=N/A \approx 0.63/r_0^2$, where $A$ denotes the size of the simulation box.

As starting configurations, we consider perfect decagonal, hexagonal, square, and random configurations. After simulations over $2 \cdot 10^{8}$ Monte-Carlo steps the energy of the final structures scatter around constant values. Typical final structures obtained for initial decagonal and random configurations are illustrated in fig. \ref{fig:overview10} for various patch widths $\sigma_{1}$. Particles are colored according to their local bond-orientational order parameter $|\psi_{10}|$. Structure factors are depicted as insets and the angular distributions are shown below each configuration.

At very narrow patch widths $\sigma_{1}=0.005$ it is difficult for the patches to find each other and the initial decagonal tiling is destroyed. At $\sigma_{1} = 0.03$ the decagonal tiling remains stable. The structure factor provides clear decagonal symmetry and the angular distribution shows sharp peaks. The bond-orientational order is maximal. Note that only particles close to contact are considered in the calculation of $|\psi_{10}|$ such that central particles appear yellow. Particles with short distances to their neighbors are stabilized by the potential minimum in combination with the correct orientation. The long characteristic distance of a decagonal quasicrystal either occurs within a pentagon with the short distance as side length or as distance between a central particle in a decagon with the short length as side length. Pentagons and decagons are supported by the favored binding angles of $\pi/5$ and $2\pi/5$. At increased patch widths bond angles deviate from the decagonal symmetry. The particles start to slightly rearrange at $\sigma_{1}=0.1$. Peaks of the angular distribution are broadened and order decreases. At $\sigma_{1}=0.2$ a dense phase with predominant distances of $r_{0}$ and angles of approximately $45^\circ$ and $66^\circ$ forms. 

Applying the given protocol with initial random configurations, particles do not arrange to decagonal quasicrystals at any patch width. At $\sigma_{1}=0.03$ most particles build a dense periodic structure. We observe angles with $72^\circ$ and $108^\circ$ that in principle could also appear in patterns with decagonal symmetry and an additional angle with $54^\circ$. Only a few elements of a decagonal tiling are found. At intermediate patch width $\sigma_{1}=0.1$ this phase competes with an Archimedean ($3^{3}4^{2}$) tiling. For initial hexagonal and square phases we obtain similar final configurations.

\begin{figure}[htb]
\centering
\includegraphics[scale=1]{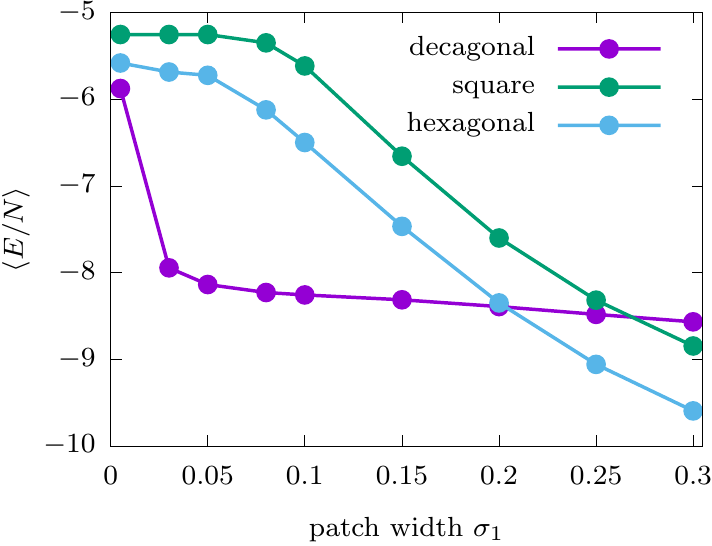} 
\caption{Average potential energy per particle of an ideal decagonal, square and hexagonal tiling as function of the patch width $\sigma_{1}$. Further potential parameters read $\sigma_{2}=0.23,\, n=2,\, r_{0}=d_{0}$ and we model colloids with 10 patches.}
\label{fig:energy10}
\end{figure}

In the following we test which structures are energetically favored by the employed potential. We apply an ideal decagonal, square and hexagonal tiling each with density $\rho \approx 0.63/r_{0}^{2}$ and depict the average potential energy per particle $\langle E/N \rangle$ as a function of the patch width $\sigma_{1}$ in fig. \ref{fig:energy10}. At low patch widths $\sigma_{1} \leq 0.2$ the ideal decagonal tiling is energetically favored. Increased patch widths allow for deviations from the decagonal angles and $\langle E/N \rangle$ decreases for all structures. At $\sigma_{1} > 0.2$ the hexagonal tiling is energetically preferred.

\begin{figure*}[htb]
\centering
\includegraphics[scale=1]{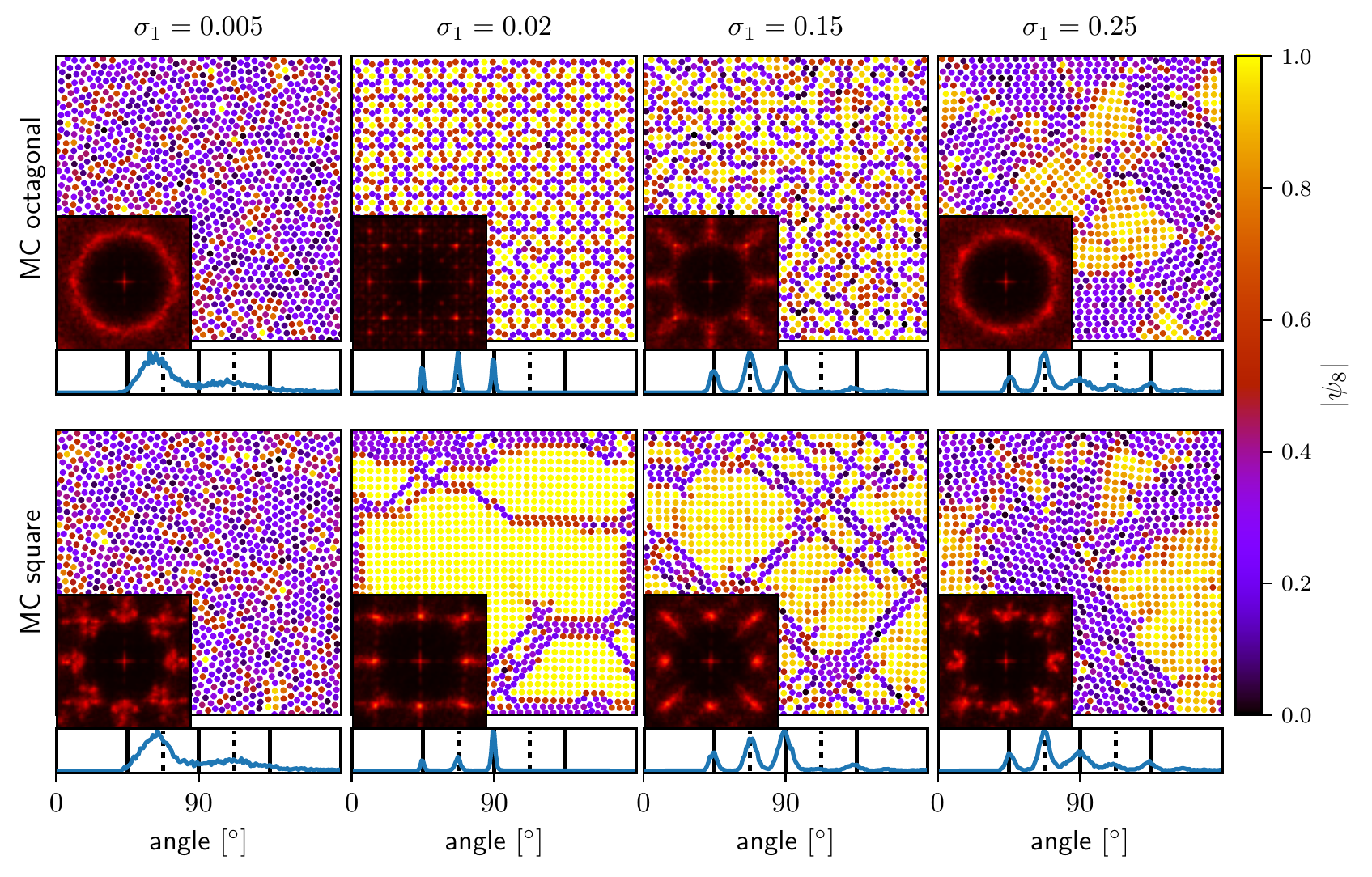} 
\caption{Configurations of $N=1393$ particles obtained from simulations with an initial octagonal Ammann-Beenker tiling (upper row) and an initial square configuration (lower row). We vary the patch width $\sigma_{1}=0.005$ (first column), $\sigma_{1}=0.02$ (second column), $\sigma_{1}=0.15$ (third column) and $\sigma_{1}=0.25$ (fourth column). Further potential parameters read $\sigma_{2}=0.29,\, n=3,\, r_{0}=l_{1}$ and we model colloids furnished with 8 patches. The color code illustrates the bond-orientational order parameter $|\psi_{8}|$. The insets depict the corresponding structure factors. The graphs below the configurations show the corresponding angular distribution functions. The black lines serve as guide to the eye and indicate ideal octagonal bond angles, i.e. $2\pi j/8, j=1,\ldots,8$. The dashed black lines indicate angles in between, i.e. $67.5^{\circ}$ and $112.5^{\circ}$. All simulations were performed at $T=0.3\epsilon$.}
\label{fig:overview8}
\end{figure*}

We now model particles with 8 symmetrically arranged patches and try to stabilize quasicrystals with octagonal symmetry. Ideal octagonal structures possess the lengths $l_{0}=2\,\mathrm{sin}(\pi/8)l_1 \approx 0.77l_1$ and $l_{1}$. In our simulations we support $l_{1}$, i.e. we chose $r_{0}=l_{1}$. In case of $r_{0}=l_{0}$ octagonal structures would be destroyed as will be shown in the next paragraph. Further potential parameters read $n=3$ and $\sigma_{2}=0.29$, while $\sigma_{1}$ is varied. Note that $\sigma_{2}$ is larger than for particles with 10 patches in order to cover the same area with patches. We adjust a particle density $\rho \approx 1.21/r_0^2$ of a perfect octagonal Ammann-Beenker tiling. Simulations are started with an ideal octagonal or a square tiling and last $5 \cdot 10^{8}$ Monte-Carlo steps. Fig. \ref{fig:overview8} depicts final configurations at $T=0.3\epsilon$. The temperature ensures mobile particles below melting.

As for particles with 10 patches the structures show low positional and orientational order at narrow patch width $\sigma_{1}=0.005$. At $\sigma_{1} = 0.02$ the octagonal quasicrystal remains stable and the angular distribution function shows sharp peaks. As for the decagonal quasicrystal, the structure is stabilized by the supported length scale in combination with the orientational part. At $\sigma_{1}=0.15$ the ideal tiling dissolves and the peaks broaden. Broad patch widths $\sigma_{1} > 0.2$ cause a coexistence of a square and zigzag phase with a predominant length $r_0$. The zigzag phase is a periodic lattice that consists of the same rhombs as in the Ammann-Beenker tiling. Zigzag and square phase both support angles of an octagonal tiling. The desired density is reached by a coexistence of both phases. 

The initial square phase does not arrange to an octagonal tiling at any patch width. Instead, parts of the square lattice rearrange into the denser zigzag phase. The number of rearranging particles increases with the patch width. Only a few local octagonal elements are found. We also perform the simulations with initial zigzag and random phases. The final structures are similar like the ones obtained from initial square phases.

\begin{figure}[htb]
\centering
\includegraphics[scale=1]{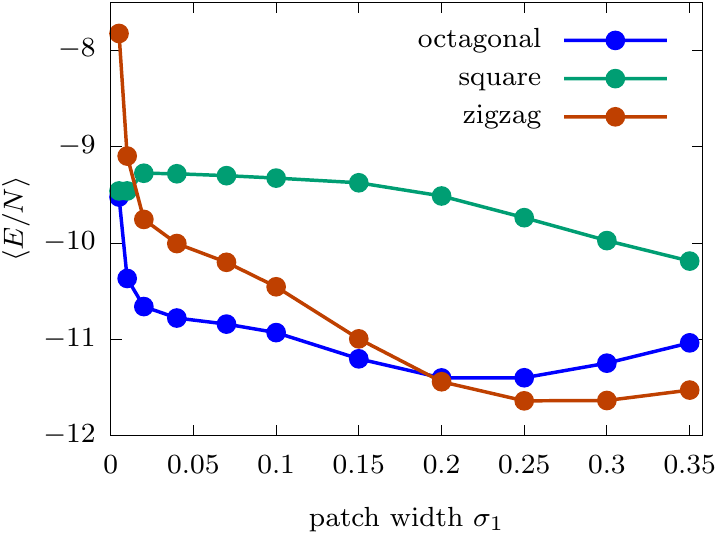} 
\caption{Average potential energy per particle of a perfect octagonal, square and zigzag tiling as function of the patch width $\sigma_{1}$. Further potential parameters read $\sigma_{2}=0.29,\, n=3,\, r_{0}=l_{1}$ and we model colloids with 8 patches.}
\label{fig:energy8}
\end{figure}

Fig. \ref{fig:energy8} illustrates the average potential energy per particle $\langle E/N \rangle$ of a perfect octagonal, square and zigzag lattice as a function of the patch width $\sigma_{1}$. All structures have the density $\rho \approx 1.21/r_{0}^{2}$ of a perfect octagonal tiling. At low patch widths $\sigma_{1} \leq 0.15$ the perfect octagonal quasicrystal is clearly energetically favored, while at larger patch widths the zigzag phase possesses the lowest energy.

\begin{figure}[htb]
\centering
\includegraphics[scale=1.1]{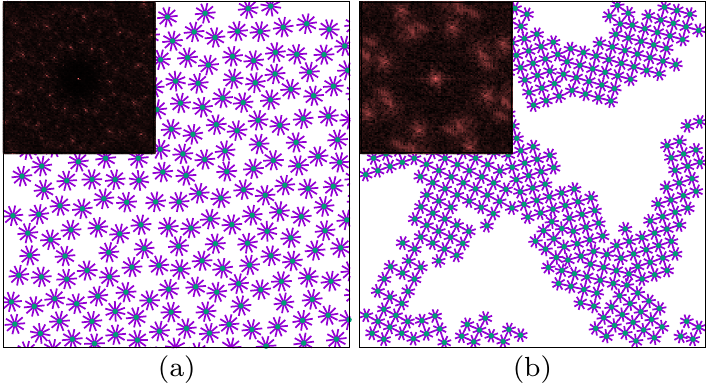} 
\caption{Variation of the potential length scale $r_{0}$. We depict configurations of particles with (a) 10 patches and supported length $r_{0}=d_{1}$ and (b) 8 patches and supported length $r_{0}= l_{0}$. The illustrated configurations are obtained from simulations with random initial states. Patches are indicated by arms around the particles. Structure factors are shown as insets.}
\label{fig:diff_length}
\end{figure}

We now study the arrangements of particles when the potential minimum is varied and another length scale is supported. To test structures with decagonal symmetry we support the long length with $r_{0}=d_1$. Further simulation parameters remain unchanged. We choose $\sigma_{1}=0.03$ for which the ideal decagonal tiling is stabilized in case of $r_{0}=d_0$. In our simulations, we do not obtain decagonal structures, i.e., neither an ideal quasicrystal remains stable, nor a random configuration self-assembles into a quasicrystal. However, in both cases the final configurations contain some decagonal motifs and pentagons. The structure factor indicates these motifs as illustrated in fig. \ref{fig:diff_length} (a). Patches are illustrated by arms around the particles. Patches of particles that are separated by the long length are oriented towards each other. Patches of particles with shorter distance do not obey the correct orientation.

In the case of particles with 8 patches we have so far supported the long length $r_0=l_1$ and now will test to support the short length, i.e. $r_{0}=l_0$. Further parameters are kept as before, and we choose $\sigma_{1}=0.02$. Independent of the initial configuration the particles arrange to dense square tilings with distances $r_{0}$ between nearest neighbor particles and with voids as shown in fig. \ref{fig:diff_length} (b) for $N=392$ particles. Note that the 8-fold symmetry suggested by the structure factor results from domains of square tilings that are rotated against each other.

\begin{figure*}[htb]
\centering
\includegraphics[scale=1]{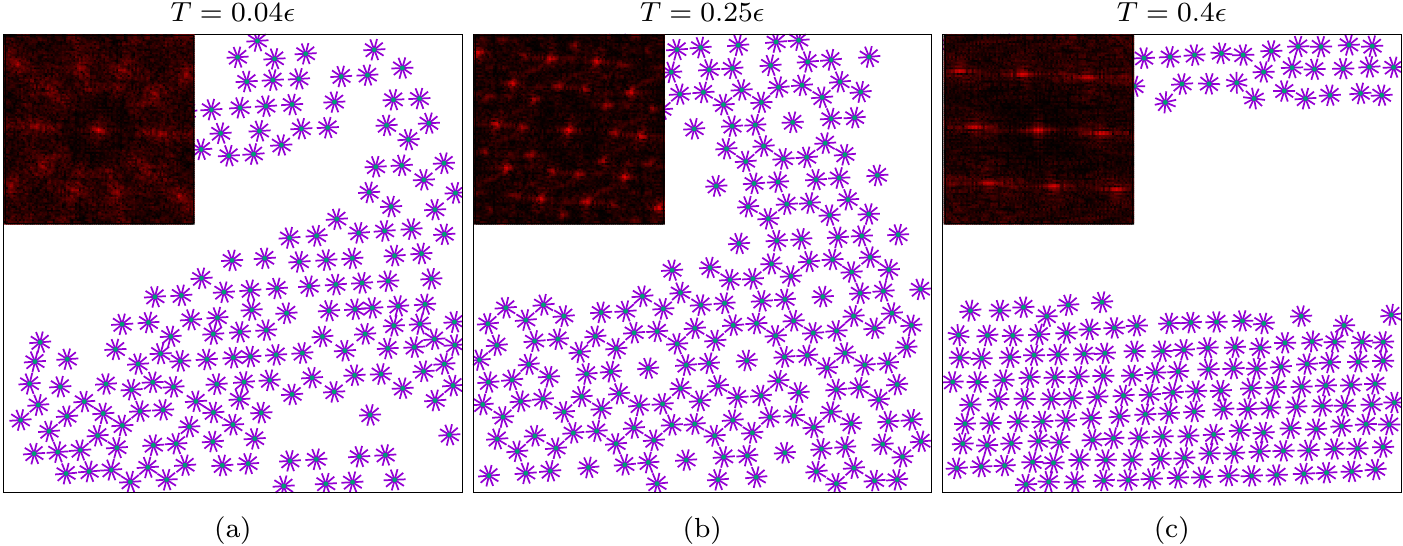} 
\caption{Configurations of $N=191$ particles with 10 symmetrically arranged patches at the surface. Patches are illustrated by arms. Starting from the fluid we simulate the system at (a) $T=0.04\epsilon$ and increase the temperature to (b) $T=0.25\epsilon$ and (c) $T=0.4\epsilon$. The potential parameters read $\sigma_{1}=0.03,\,\sigma_{2}=0.23,\, n=2,\, r_{0}=d_{0}$. The insets show the structure factors of the corresponding configurations.}
\label{fig:nice_5fold}
\end{figure*}

So far we have found metastable octagonal or decagonal quasicrystals. These structures can be observed if the simulations are started with the respective structures. In the following we check if we can obtain quasicrystals with octagonal or decagonal symmetry from a random initial state. We first model colloids furnished with 10 patches. We choose the same potential parameters as in previous simulations and stabilize the short length $d_{0}$. Simulations with a low temperature $T=0.04\epsilon$ result in elements of a decagonal tiling (see fig. \ref{fig:nice_5fold} (a)) and indications of a decagonal symmetry provided by the structure factor (inset of fig. \ref{fig:nice_5fold} (a)). We increase the temperature of this configuration to $T=0.25\epsilon$. Beside conventional displacement and rotation moves we propose additional jumps that can correspond to phasonic flips, i.e. rearrangements within quasicrystals that do not change or at least almost not change the total energy of the structure \cite{Kromer,Sandbrink,Martinsons,Hielscher}. Such flips are implemented by proposing displacements in a random direction with the distance $d_{\mathrm{flip}}=d_{1}-d_{0}$ at which flips usually occur. We propose such possible phasonic flip steps with a probability of $p=0.2$. We obtain a quasicrystalline arrangement of particles as illustrated in fig. \ref{fig:nice_5fold} (b). The patches indicated by arms around the particles are nicely oriented towards each other. The long length scale is supported indirectly by the energy gain provided by the angular term. However, we observe an excess of the short length compared to the ideal tiling and voids arise. The structure is metastable and increased temperatures lead to a transition to a triangular phase as depicted for $T=0.4\epsilon$ in fig. \ref{fig:nice_5fold} (c). 

In case of colloids with 8 patches we also apply the potential parameters as in previous investigations and stabilize the long length $l_{1}$. Even for low temperatures the particles arrange to a square tiling. Patches are oriented towards the patches of nearest and next nearest neighbors. Since square tilings are also supported by 8 patches, it is hard to build an octagonal quasicrystal. For temperatures $T < 10^{-4}$ we obtain random frozen arrangements that again turn into square tilings at increased temperatures. 

\section{Conclusions}

We have studied how quasicrystals can be stabilized in patchy colloidal systems in two dimensions. Even though only one typical length can be supported by the potential minimum, the quasicrystalline order with two incommensurate lengths is induced by an additional anisotropic contribution. While it is known that dodecagonal orderings can be easily found even in systems in which the number of patches seems not to fit to the 12-fold symmetry \cite{vdLinden,Reinhardt,Gemeinhardt}, octagonal and decagonal quasicrystals seem to be harder to realize in soft matter systems. Here we have shown that one needs narrow patches corresponding to sharply enforced preferred binding angles in order to stabilize quasicrystals with 8- or 10-fold symmetry. We have shown that these structures are energetically favored in case of narrow patches and that they can stay stable in Monte-Carlo simulations that have been started with the respective patterns.

We have also shown that decagonal quasicrystals can be obtained from random initial conditions by cooling and subsequent reheating. Therefore, there are protocols that can be used to receive these quasicrystals as metastable structures. However, it remains an open question whether octagonal or decagonal structures might exist in thermal equilibrium in systems with patchy colloids (maybe with modified potentials).
 
Our results contribute to a better understanding of why metallic quasicrystals do not possess dodecagonal symmetry as often as soft matter quasicrystals \cite{Dotera}. The reason might be due to preferred binding angles that are usually given in a very sharp way in case of metallic systems. Note that our findings are consistent with experimental results of quasicrystals in metallic alloys. Most stable metallic quasicrystals possess icosahedral symmetry. Also a few decagonal phases have been observed, while octagonal structures provide least examples and are usually metastable \cite{Steurer}.

\acknowledgments

The work was supported by the German Research Foundation (DFG) by grant Schm 2657/4.

\end{document}